\documentclass[twoside,leqno,twocolumn]{article}

\usepackage[letterpaper]{geometry}
\usepackage{ltexpprt}
\usepackage{hyperref}

\usepackage{cite}
\usepackage{amsmath,amssymb,amsfonts}
\usepackage{algorithm}
\usepackage{algpseudocode}
\usepackage{graphicx}
\usepackage{pifont}
\usepackage{textcomp}
\usepackage{xcolor}
\usepackage{subfigure}
\usepackage{multirow}
\usepackage{booktabs}
\usepackage[table]{xcolor}
\usepackage{multirow}
\usepackage{makecell}
\usepackage{enumitem}
\usepackage{arydshln}

\definecolor{myblue}{RGB}{48,60,203}

\definecolor{myred}{RGB}{220,79,62}

\newcolumntype{T}{>{\fontsize{7.5}{10}\selectfont}c}

\def\BibTeX{{\rm B\kern-.05em{\sc i\kern-.025em b}\kern-.08em
    T\kern-.1667em\lower.7ex\hbox{E}\kern-.125emX}}

\begin{document}

\newcommand\relatedversion{}
\newcommand\keywords[1]{%
  \vspace{1em}\noindent\textbf{keywords: }#1}

\title{\Large Towards Hierarchical Multi-Agent Decision-Making for Uncertainty-Aware EV Charging}

\author{
Lo Pang-Yun Ting\textsuperscript{1}\textsuperscript{2}\thanks{Work done during Lo Pang-Yun Ting being a visiting scholar at Arizona State University.}, 
Ali Şenol\textsuperscript{2}, 
Huan-Yang Wang\textsuperscript{1},
Hsu-Chao Lai\textsuperscript{1},\\
Kun-Ta Chuang\textsuperscript{1},
Huan Liu\textsuperscript{2}\\[1ex]
\small $^{1}$ Dept. of Computer Science and Information Engineering, National Cheng Kung University, Taiwan \\
\small Emails: \{lpyting, hywang, hclai\}@netdb.csie.ncku.edu.tw, ktchuang@mail.ncku.edu.tw \\
\small $^{2}$ School of Computing and Augmented Intelligence, Arizona State University, USA\\
\small Emails: \{lting5, asenol, huanliu\}@asu.edu \\
}

\date{}

\maketitle

\begin{abstract}

Recent advances in bidirectional EV charging and discharging systems have spurred interest in workplace applications. However, real-world deployments face various dynamic factors, such as fluctuating electricity prices and uncertain EV departure times, that hinder effective energy management. To address these issues and minimize building electricity costs while meeting EV charging requirements, we design a hierarchical multi-agent structure in which a high-level agent coordinates overall charge or discharge decisions based on real-time pricing, while multiple low-level agents manage individual power level accordingly. For uncertain EV departure times, we propose a novel uncertainty-aware critic augmentation mechanism for low-level agents that improves the evaluation of power-level decisions and ensures robust control under such uncertainty. Building upon these two key designs, we introduce \emph{HUCA}, a real-time charging control framework that coordinates energy supply among the building and EVs. Experiments on real-world electricity datasets show that \emph{HUCA} significantly reduces electricity costs and maintains competitive performance in meeting EV charging requirements under both simulated \textit{certain and uncertain departure scenarios}. The results further highlight the importance of hierarchical control and the proposed critic augmentation under the uncertain departure scenario. A case study illustrates \emph{HUCA}'s capability to allocate energy between the building and EVs in real time, underscoring its potential for practical use.

\end{abstract}


\section{Introduction}

\begin{figure}[t]
\graphicspath{{figs/}}
\begin{center}
\vspace{-2mm}
\includegraphics[width=0.48\textwidth]{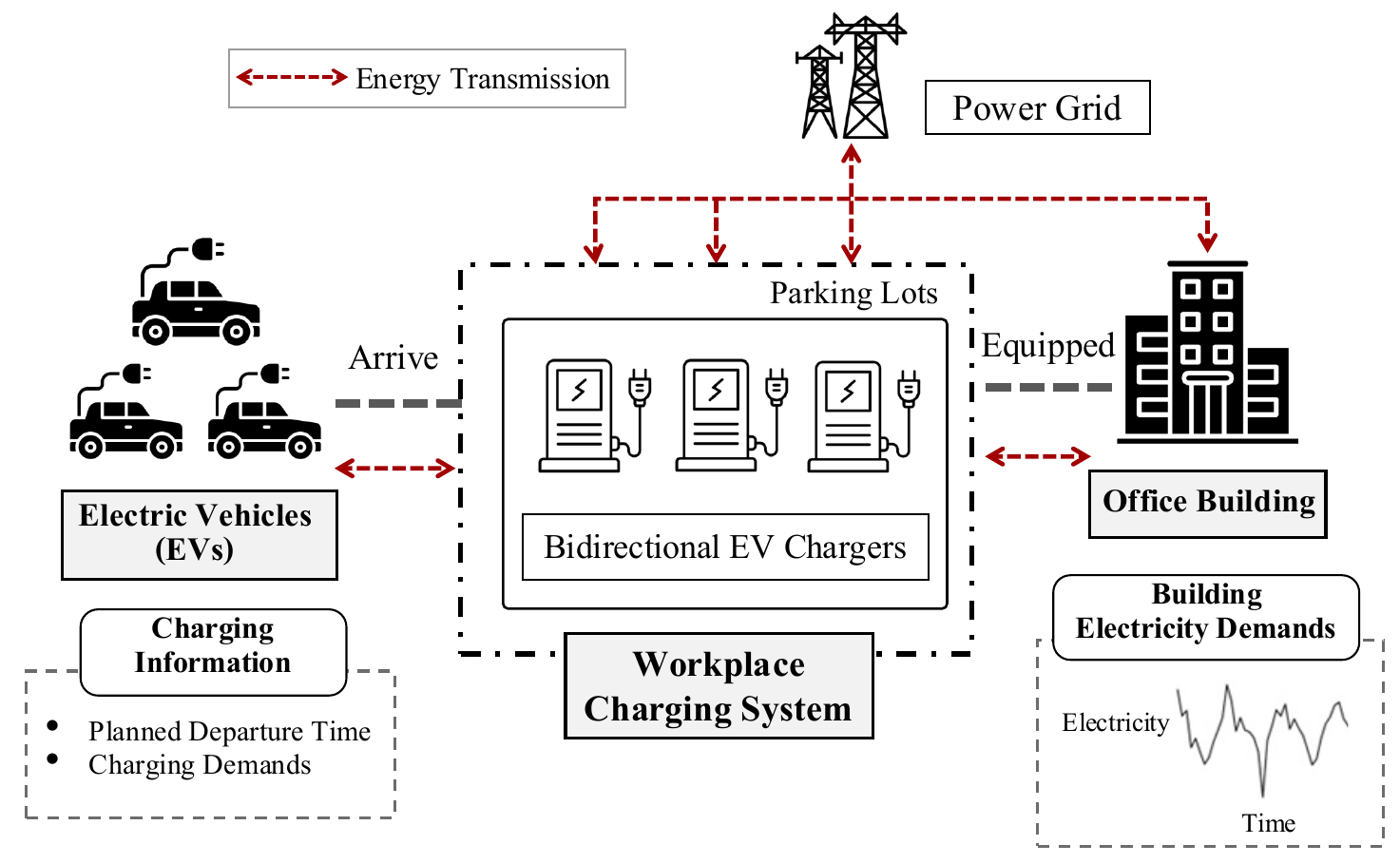}
\end{center}
\vspace{-1em}
\caption{The illustration of the workplace charging scenario with the bidirectional EV chargers.}
\vspace{-4.8mm}
\label{fig:system_overview}
\end{figure}

\noindent\textbf{Background.} The rapid growth in Electric Vehicle (EV) adoption, driven by global sustainability efforts, has led to a surge in demand for advanced charging solutions.
According to a report by CBRE Group (Coldwell Banker Richard Ellis Group) Incorporated~\cite{ev.demandFroChargingPile}, workplace charging sessions increased at twice the rate of new charging station installations in 2023. This highlights the growing interest for EV charging options in the workplace. Meanwhile, recent advances in \textit{bidirectional charging systems}~\cite{sami2019bidirectional}~\cite{upputuri2023comprehensive}, which support both Grid-to-Vehicle (G2V) and Vehicle-to-Grid (V2G) power flows, have shown the potential to enhance flexibility with regard to charging stability and response capabilities.
The effectiveness of bidirectional charging systems has inspired us to explore how better scheduling of G2V and V2G timings can balance the demand for EV charging and reduce overall electricity costs. 

In particular, EVs often remain stationary for long periods at office locations, as expected. This prolonged parking time allows for more flexible charging strategies, enabling energy transfer between EVs and office building. An example of a workplace charging system is shown in Figure~\ref{fig:system_overview}.
Upon arrival, each EV is coordinated by the charging system, with users first providing information such as their charging requirements and anticipated departure time. With such information, the system decides whether to discharge energy from the EV to the building (using V2G) during high-price periods or to charge the EV from the grid to meet its demand before departure (using G2V). Either the V2G or the G2V option can be determined on-the-fly according to the optimal decision-making criteria.

\noindent\textbf{Challenges.} For real-time charging control of EVs in various scenarios, previous studies~\cite{saner2022cooperative}~\cite{Silva2020CoordinationOE} have explored the use of multi-agent reinforcement learning (MARL) techniques to regulate EV charging actions. However, most existing approaches fail to consider real-world \textit{\underline{dynamic factors}}, such as dynamic energy prices and the possibility that EV users may depart earlier than the expected time, which complicate determining optimal control strategies for each EV. Moreover, to avert transformer overloads\footnote{Transformer overloads occur when the demand for electricity exceeds the capacity of a transformer, thereby increasing the risk of overheating and subsequent failure.} that could destabilize the power grid~\cite{visakh2023analysis}, it is necessary to impose \textit{\underline{charging power limits}}, thereby further complicating the management of EV charging. These dynamics and limitations pose significant challenges in balancing the energy supply between the building and EVs while minimizing electricity costs. It is crucial to recognize that managing charging improperly could result in considerably higher electricity bills, as power companies will levy extra charges due to overconsumption of energy~\cite{rosado2020framework}.

\noindent\textbf{Proposed Method.}
To tackle these challenges, we propose \emph{\textbf{HUCA}} (\textbf{\underline{H}}ierarchical Multi-Agent Control with \textbf{\underline{U}}ncertainty-Aware \textbf{\underline{C}}ritic \textbf{\underline{A}}ugmentation), a novel framework designed for real-time charging control. \emph{HUCA} operates under dynamic pricing and penalty mechanisms, consistent with practical situations. A new hierarchical multi-agent framework is devised to balance the energy supply between the building and EVs in dynamic and uncertain environments. \emph{HUCA} consists of two levels of control: a high-level agent and multiple low-level agents. The \textbf{high-level agent} determines whether to charge or discharge EVs, which is a collective decision for all EVs. Based on the high-level decision, the \textbf{low-level agents} collaboratively modulate the individual charging (or discharging) power level for each EV, ensuring they stay within the specified charging power limits, with the goal of satisfying anticipated demands and avoiding transformer overload.

A key innovation of \emph{HUCA} lies in its capability to handle EVs that deviate from their expected departure times. These unpredictable departures introduce uncertainty, potentially disrupting low-level agents' charging decisions. To address this, we propose an \textbf{uncertainty-aware critic augmentation} mechanism, which assesses the likelihood of departure deviation and adjusts the assessments of low-level agents' charging decisions accordingly. This mechanism achieves two objectives: \underline{\textbf{(\textit{i})}} accounting for uncertainties into the assessment of low-level agents' decisions, and \underline{\textbf{(\textit{ii})}} limiting the direct effect of uncertainties to the present time slot, ensuring the robustness of future decision assessments. Therefore, \emph{HUCA} dynamically adapts to arbitrary EV behaviors while maintaining robust control. \emph{HUCA} is designed to minimize the electricity costs of the building while striving to fulfill EV charging demands in a dynamic environment. 

The main contributions in this paper are as follows:

\begin{itemize}[leftmargin=*]
    \item \underline{\textbf{Real-Time Charging Control}}: We propose \emph{HUCA}, a novel hierarchical multi-agent structure that balances energy supply between the building and EVs, optimizing charging controls on-the-fly.
    \item \underline{\textbf{Uncertainties and Limitations Handling}}: High-level and low-level agents  determine optimal actions under dynamic pricing and power stability constraints. In addition, an uncertainty-aware augmentation is designed to adaptively adjust decisions to handle uncertain EV departures.
    \item \underline{\textbf{Practical Effectiveness}}: 
    We evaluate \emph{HUCA} through analysis of real-world datasets, incorporating simulated EV behaviors in scenarios with both certain and uncertain departure times.
    Results demonstrate that \emph{HUCA} achieves the lowest electricity costs while maintaining competitive performance in fulfilling EV charging requirements, which shows its potential as an AI-driven solution for intelligent vehicle charging control.
\end{itemize}

\section{Related Work}

\subsection{Real-time Charging Control.}

Optimizing charging control considers different pricing mechanisms, such as time-of-use pricing~\cite{ting2024online, Zhang2023ElectricVL}, real-time pricing~\cite{bitencourt2017optimal, selim2021electric}, and so on. Also, Day-ahead optimization combined with real-time control has been used to address uncertainties in solar output and EV parking behaviors~\cite{Guo2016TwoStageEO}, electricity prices~\cite{Zhao2017RiskBasedDS} as well as energy demand and PV generation~\cite{Yan2019OptimizedOC}. Integrated management systems, additional energy resources or battery energy storage systems have also been developed for the adaptive control~\cite{Chen2020PowerCS, Chopra2023CostBA, ChandraMouli2019IntegratedPC}.

\subsection{Reinforcement Learning for Charging Management.}

Single-agent RL, multi-agent RL (MARL), and hierarchical MARL approaches were used to manage charging systems. Single-agent RL represented the control system~\cite{Li2020ConstrainedEC}~\cite{Mhaisen2020RealTimeSF} under specific considerations, such as ensuring EV battery requirements~\cite{Zhang2021CDDPGAD}, addressing charger shortage~\cite{Li2023OptimalEC}, maximizing profits for charging stations~\cite{wang2019reinforcement}, and minimizing users' charging costs~\cite{Yan2021DeepRL}. MARL modeled fine-grained cooperations~\cite{Silva2020CoordinationOE, Li2024MultiAgentGR, Park2022MultiagentDR, Yan2022ACC} among EVs and buildings to tackle safety concerns, power transfer overload~\cite{Silva2020CoordinationOE}, and users' charging cost or operating cost~\cite{Li2024MultiAgentGR, Park2022MultiagentDR, Yan2022ACC}. Hierarchical MARL further separated system operators and EV users, and coordinated them to minimize demand charges and energy costs~\cite{saner2022cooperative}.

Although these studies optimized charging control with various aspects, most studies do not address the issue of uncertain EV departures. While some works~\cite{Mhaisen2020RealTimeSF, Guo2016TwoStageEO, Park2022MultiagentDR, wang2019reinforcement} consider uncertainties in EV departure, they primarily focused on a single objective, such as maximizing system revenue or minimizing user charging costs.

Furthermore, in many of these studies, uncertainty factors are only encoded as part of the state information or leave them implicit in the stochastic transition dynamics in their RL models, \textbf{without explicitly modeling or quantifying it}, which limits the model’s ability to adaptively adjust charging control under varying levels of departure uncertainty.

In contrast, our work addresses dynamic factors, including fluctuating electricity prices, EV charging demands, and the energy requirements of the office building. More importantly, we explicitly incorporate and quantify EV departure uncertainty in our model, allowing it to directly shape the reinforcement learning decision-making process, while avoiding current uncertainty from misleading future charging decisions. This enables reliable charging control that simultaneously minimizes total electricity costs and accounts for EV users' charging demands in a dynamic environment.

\section{Preliminaries}

\begin{figure*}[h]
\graphicspath{{figs/}}
\begin{center}
\includegraphics[width=0.96\textwidth]{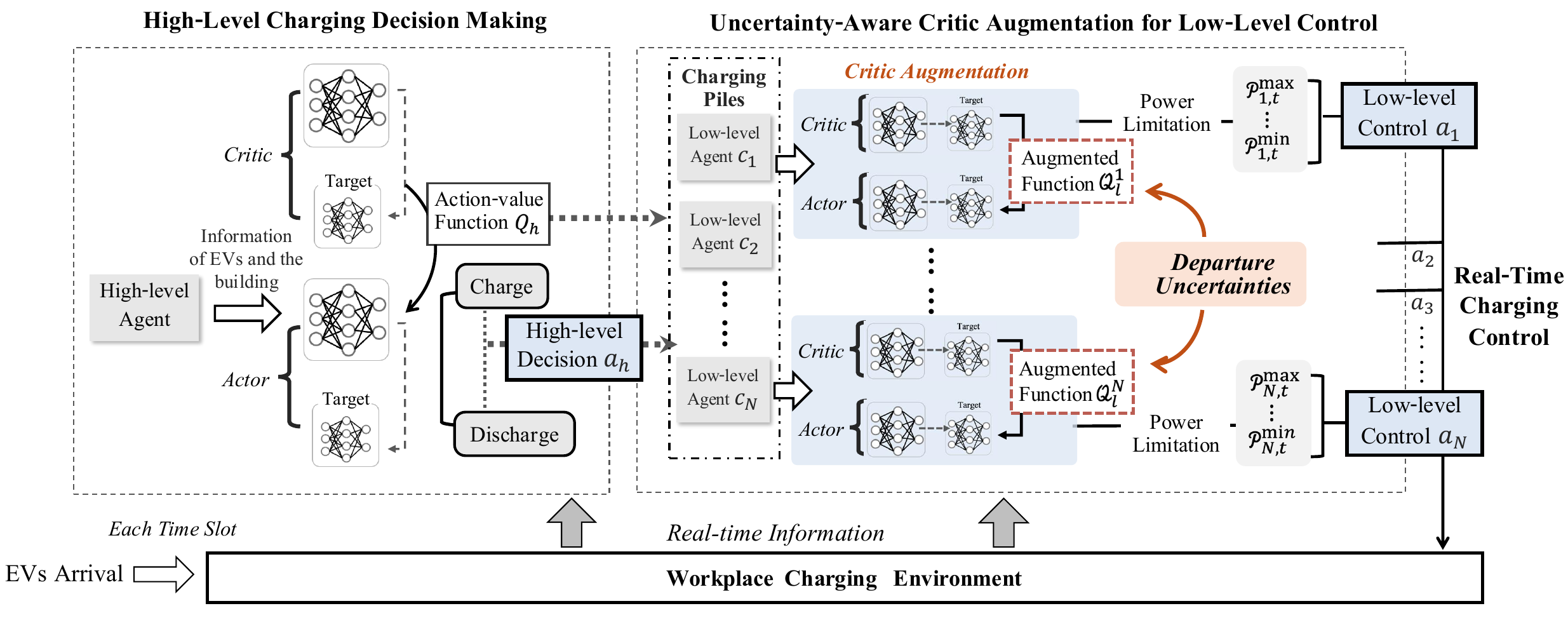}
\end{center}
\vspace{-0.8em}
\caption{The overview of the \emph{HUCA} framework. The high-level agent decides to charge or discharge EVs, while low-level agents control the power of each charging pile within power limitations.}
\label{fig:framework}
\end{figure*}

We describe key symbols, definitions, and general formulations of the Markov decision process (MDP) and deep deterministic policy gradient (DDPG) in reinforcement learning.

\subsection{Key Symbols and Definitions.}
\label{subsec:definition}

Let $\mathcal{C}=\{c_1, c_2, ..., c_N\}$ denote a set of charging piles (abbreviated as piles henceforth) located at the office building's charging station. $\mathcal{V}_t=\{v_1, v_2, ..., v_M\}$ specifies a set of EVs docking at piles at time slot $t$, where $M\le N$. The charging capacity of the charging station is denoted as $\mathcal{P}^{\text{max}}$ (kW), which represents the maximum allowable charging power at the station ($\mathcal{P}^{\text{max}}>0$).

\noindent\textbf{Definition 1. (State of Charge (SoC)):} The state of charge (SoC) of an EV battery represents the percentage of energy stored in the battery. For each EV $v_i \in \mathcal{V}_t$, the SoC of $v_i$ at time slot $t$ is denoted as $SoC^i_t \in [0\%, 100\%]$.

\noindent\textbf{Definition 2. (EV Charging Information):} In our scenario, when an EV arrives at the workplace charging system (Fig.~\ref{fig:system_overview}) at time slot $t$ and connects to pile $c_i$, the EV user sends the charging information $\mathcal{I}^i$ to the charging system. The charging information is denoted as a tuple: $\mathcal{I}^i = (t^i_\text{arr}, t^i_\text{dep}, SoC^i_\text{arr}, SoC^i_\text{dep}, C^i)$, where $t^i_\text{arr}$ and $t^i_\text{dep}$ represent the arrival time and the \textit{planned} departure time, respectively. $SoC^i_\text{arr}$ and $SoC^i_\text{dep}$ denote the SoC of the EV battery upon arrival and the \textit{expected} SoC to be reached by the charging system before the departure time, respectively. $C^i$ is the battery capacity of the EV.
Note that in our scenario, the actual departure time of the EV user may randomly occur earlier than the planned time $t^i_\text{dep}$.

\noindent\textbf{Definition 3. (Charging Power Limitation):} To ensure power system stability, the charging power of each charging pile is limited by the maximum charging capacity $\mathcal{P}^{\text{max}}$ of the charging station, following the settings in ~\cite{Li2023OptimalEC}~\cite{li2023constrained}. Given the number of EVs $|\mathcal{V}_t|$ docking at the charging station at time slot $t$, the maximum charging power $P^{\text{pile}}_t$ of each pile at time slot $t$ is estimated as $P^{\text{pile}}_t = \mathcal{P}^{\text{max}}/|\mathcal{V}_t|$. The maximum discharging power of each pile is $-P^{\text{pile}}_t$.

To consider the fulfillment of the expected SoC $SoC^j_{\text{dep}}$ of each EV $v_j$, we estimate the specific charging/discharging power boundary that pile $c_i$ provides to the connected EV $v_j$ at each time slot $t$. Specifically, given the minimum and maximum SoC that EV $v_j$ can reach at time $t$, denoted by $SoC^{LB}_{t}$ and $SoC^{UB}_{t}$, respectively, the specific charging and discharging power boundaries of pile $c_i$ are formulated according to~\cite{Li2023OptimalEC}\cite{li2023constrained} as:

\begin{equation*}
\label{eq:app_boundary}
\left\{\begin{matrix}
\mathcal{P}^{\max}_{i,t}=\min\bigl(P^{\text{pile}}_t,(SoC^{UB}_{t}-SoC^j_{t-1}\cdot C^j\cdot\eta^*)\bigr) \\
\mathcal{P}^{\min}_{i,t}=\max\bigl(-P^{\text{pile}}_t,(SoC^{LB}_{t}-SoC^j_{t-1}\cdot C^j\cdot\eta^*)\bigr)
\end{matrix}\right.,
\vspace{-0.5em}
\end{equation*}
where $SoC^j_{t-1}$ is the SoC of $v_j$ at time $t-1$, and $C^j$ denotes the battery capacity for $v_j$. $\eta^*$ is function of charging efficiency.

\subsection{Markov Decision Process and Deep Deterministic Policy Gradient.}
Decision-making problems are commonly modeled as MDPs, defined by the four-tuple information $(S,A,R,T)$: states $S$, actions $A$, rewards $R$, and the state transition probabilities $T$. The main objective of an MDP is to find a policy that maximizes the cumulative rewards by selecting appropriate actions. The Deep Deterministic Policy Gradient (DDPG) algorithm~\cite{lillicrap2015continuous}, which is an actor-critic policy-based method that extends stochastic policy gradients to deterministic settings, has shown great effectiveness in complex environments. Specifically, its deep actor network, parameterized by $\theta^{\mu}$, approximates the deterministic policy $\mu_{\theta}: S\to  A$. The critic network, parameterized by $\theta^{Q}$, estimates the action-value function $Q(s,a|\theta^{Q})$. Overall, the actor network determines the optimal action for a given state, while the critic network evaluates the action's quality.

\subsection{Problem Formulation.}

Our work aims to benefit both the office building and EVs by enhancing the original single-agent DDPG framework. We model the problem as a hierarchical MDP to optimize multiple objectives simultaneously, aiming to find the most effective bidirectional charging strategy that reduces building overall electricity costs  while considering EV charging needs. At each time slot $t$, given the charging pile set $\mathcal{C}$, EV set $\mathcal{V}_t$, EV charging information $\{\mathcal{I}^j|v_j \in \mathcal{V}_t\}$, and the charging and discharging power boundaries $\{\mathcal{P}^{\max}_{i,t}, \mathcal{P}^{\min}_{i,t} | c_i \in \mathcal{C}\}$, the objective is to determine the optimal charging (or discharging) power $\mathcal{P}^{\text{opt}}_{i,t}$ for the EV connected to pile $c_i$ at time slot $t$, where $ \mathcal{P}^{\min}_{i,t} \le \mathcal{P}^{\text{opt}}_{i,t} \le \mathcal{P}^{\max}_{i,t}$.

By determining $\mathcal{P}^{\text{opt}}_{i,t}$, in the \textbf{certain departure scenario}, where the EV's actual departure time matches the planned one, the goal is to minimize the total electricity cost of the building while ensuring that EV users' expected SoCs are met at their departure. In \textbf{uncertain departure scenario}, where EVs depart earlier than expected\footnote{When the departure occurs later than anticipated, achieving desired SoCs of these EVs is relatively straightforward compared to scenarios of early departure. Consequently, this paper concentrates on addressing the uncertainty of early departure.}, our objective is to achieve their desired SoCs to the greatest extent while minimizing costs, balancing the trade-off between cost reduction and maximizing EV charging fulfillment.

\section{The \emph{HUCA} Framework}

The architecture of \emph{HUCA} is illustrated in Figure~\ref{fig:framework}. Initially, a high-level agent determines whether to charge or discharge EVs using real-time data (Sec.~\ref{subsec:high-level}). Subsequently, multiple low-level agents manage the charging and discharging power for each individual pile, taking into account the unpredictability of EV departures (Sec.~\ref{subsec:low-level-uncertain-aug}). Finally, \emph{HUCA} can determine the optimal decision instantaneously (Sec.~\ref{subsec:optimal_strategy}).

\subsection{High-Level Charging Decision Making.}
\label{subsec:high-level}

This section formulates a novel MDP for a high-level agent to decide whether to charge or discharge EVs.

\noindent \textbf{State representations.} A state $s_h$ represents the status at time slot $t$, consisting of three main information $s_h=(I^{e}_t, I^{v}_t, I^l_{t-1})$:

\begin{itemize}
    \item $I^{e}_t$ includes the electricity load of the building and electricity price for the following periods: the current time slot, the average over the past $n$ hours, and the historical price at the same time and day of the week.
    \item $I^{v}_t$ includes the current number of EVs at the charging station and the amount of energy provided to them.
    \item $I^{l}_{t-1}$ includes the average and the standard deviation of critic value of all low-level agents from at time $t-1$.
\end{itemize}

\noindent \textbf{Discrete actions for the high-level agent.} Let $a_h \in [0,1]$ denote the action selected by the high-level agent. It is then converted into the discrete high-level action $a^{\text{disc}}_h$ to determine whether to charge or discharge EVs, as defined below:

\begin{equation}
\label{eq:high_level_action}
a^{\text{disc}}_h=\left\{\begin{matrix}
1  & \text{, if $a_h\ge 0.5$}  \\
0  & \text{, if $a_h < 0.5$}
\end{matrix}\right.,
\end{equation}
where $a^{\text{disc}}_h=1$ represents charging, and $a^{\text{disc}}_h=0$ denotes discharging.

\noindent \textbf{High-level objective and reward function.}
The total electricity cost of a building is calculated based on two parts: (i) the total electricity consumption and (ii) the amount of electricity exceeding the contracted capacity\footnote{Typically, the electricity provider establishes a maximum power limit for each subscriber to efficiently manage overall consumption. If the instant electricity consumption of a subscriber surpasses this agreed-upon capacity, they will incur additional penalties.}~\cite{Fernndez2013CostOO}. The objective of the high-level agent is to minimize the total electricity cost. 

Let $L_t$ represent the electricity load, $\Delta L_{t}$ denote the excess electricity consumption of the building, $p_{t}$ be the electricity price, and $t$ specify the time slot. Define $[L_{t'}]_{+}=\max(0, \Delta L_{t'})$. The reward function, $r_h(s_h, a^{\text{disc}}_h)$, abbreviated as $r_h$, is formulated as:

\vspace{-1em}
\begin{equation}
\label{eq:high_level_reward}
\begin{aligned}
&\hspace*{-0.1cm}
r_h
=\kappa \cdot \Bigl(\underbrace{-\sum_{t'=1}^{t} p_{t'} \cdot L_{t'} -   [L_{t'}]_{+} \cdot \varphi }_{\text{electricity cost term}} \Bigr) 
\\
&\hspace*{0.4cm}
+ \underbrace{\Bigl( -|L_{t} - L_{\text{avg}}| \Bigl)}_{\text{electricity balance term}},
\end{aligned}
\end{equation}
where the first term represents the potential total electricity cost up to time $t$, weighted by an importance factor $\kappa \in [0,1]$. $\varphi \in \mathbb{R}$ is a fixed penalty coefficient weighting the exceeding instant electricity usage. The second term balances the electricity consumption compared to the previous average load $L_{\text{avg}}$. This design \underline{reduces the reward for costly} or \underline{imbalanced charging}, encouraging the high-level agent to avoid similar actions in future comparable states.

To evaluate charging or discharging actions, we apply DDPG concepts to update the high-level policy $\mu_h$. The high-level agent trains the critic network, with parameters $\theta^{Q}_h$, to approximate action-value function  $Q_h(s_h,a_h)$ by minimizing the following loss:

\vspace{-0.6em}
\begin{equation}
\label{eq:high_level_critic_loss}
{\small	
\begin{aligned}
&\hspace*{-0.1cm}
\mathcal{L}(\theta^{Q}_h)=\mathop{\mathbb{E}}_{s_h,a_h,r_h,s'_h\sim \mathcal{D}_h}\biggl[\Bigl(Q_h(s_h, a_h)-y_h\Bigr)^2\biggr],
\\
&\hspace*{-0.1cm}
y_h=r_h+\gamma\bar{Q}_h(s'_h,a'_h)|_{a'_h=\bar{\mu}_h(s'_h)},
\end{aligned}
}
\end{equation}
where $\gamma$ is a discount factor. $s'_h$ is the next state after taking action $a_h$. $\bar{Q}_h$ and $\bar{\mu}_h$ are the target action-value function and target policy, respectively, used to stabilize training. The replay buffer $\mathcal{D}_h$ stores the transition experiences of the high-level agent in the form of tuples $(s_h, a_h, r_h, s'_h)$. Subsequently, the actor network, with parameters $\theta^{\mu}_h$, refines the high-level policy $\mu_h$ via gradient descent as follows:

\begin{equation}
\label{eq:high_level_actor_loss}
\nabla _{\theta^{\mu}_h}J(\mu_{h})\!=\!\!\!\!\!\!\mathop{\mathbb{E}}_{s_h\sim\mathcal{D}_h}\!\nabla _{\theta^{\mu}_h}\mu_{h}(a_h|s_h)\nabla _{a_h}Q_{h}(s_h,a_h)|_{a_h=\mu_{h}(s_h)}.
\end{equation}

Using mini-batch updates, the high-level agent’s training loss is computed over transition experiences drawn from the replay buffer $\mathcal{D}_h$. At each training step, given a mini-batch $\mathcal{B}_h$ of transitions $\bigl\{(s_h^{(b)}, a_h^{(b)}, r_h^{(b)}, s_h^{(b)})\bigr\}_{b=1}^{|\mathcal{B}_h|}$ drawn from $\mathcal{D}_h$, the critic and actor networks, parameterized by $\theta^Q_h$ and $\theta^{\mu}_h$, are updated as shown in Algorithm~\ref{algo:high_update}.

\begin{algorithm}[t]
\small
\caption{High-Level Agent Update}
\label{algo:high_update}
\begin{algorithmic}[1]
\Procedure{UpdateHigh}{$\mathcal{B}_h$, $\theta^Q_h$, $\theta^{\mu}_h$}
\State Update the critic network by minimizing:

$\mathcal{L}(\theta^Q_h) = \frac{1}{\mathcal{B}_h} \sum_b \left(Q_h(s^{(b)}_h, a^{(b)}_h) - y{(b)}_h\right)^2
$
\State Update the actor netwok using:

$\nabla_{\theta^\mu_h} J \approx \frac{1}{\mathcal{B}_h} \sum_b \nabla_{\theta^\mu_h} \mu_h(s^{(b)}_h) \nabla_{a_h} Q_h(s^{(b)}_h, a^{(b)}_h)$
\EndProcedure
\end{algorithmic}
\end{algorithm}

\subsection{Low-Level Control with Uncertainty-Aware Critic Augmentation.}
\label{subsec:low-level-uncertain-aug}

Based on the charging or discharging decision made by the high-level action, the goal of the low-level control is to determine the optimal charging (or discharging) power level for each pile, taking into account the uncertainty of EV departures. Each pile $c_i \in \mathcal{C}$ is considered as a low-level agent and its task is formulated as an MDP. A multi-agent structure including multiple piles is deployed for low-level control.

\noindent \textbf{State representations.} For a low-level agent $c_i \in \mathcal{C}$, the state $s^i_l$ observed at time slot $t$ can be described as comprising two main information $s^i_l=(I^{v_j}_t, I^h_t)$. Details are:

\begin{itemize}
    \item $I^{v_j}_t$ includes the current SoC of EV $v_j$ and the suitable maximum and minimum charging powers ($\mathcal{P}^{\max}_{i, t}$ and $\mathcal{P}^{\min}_{i,t}$) of agent (charging pile) $c_i$ as defined in Eq.~\ref{eq:app_boundary}.
    \item $I^h_t$ includes the discrete high-level action $a^{\text{disc}}_h$ determined at time slot $t$, and the critic value of state-action pair $(s_h, a_h)$ obtained from the high-level critic network $Q_h(s_h, a_h; \theta^Q_h)$ at time slot $t$.
\end{itemize}

\subsubsection{\textbf{Low-level objective and the critic augmentation.}}

Multi-agent DDPG (MADDPG)~\cite{lowe2017multi} extends DDPG into a multi-agent policy gradient algorithm, where decentralized agents learn a centralized critic based on the observations and actions of all agents. Inspired by MADDPG, we formulate the objective of each critic network, $\theta^{Q,i}_h$, of a low-level agent $c_i \in \mathcal{C}$ by minimizing the following loss:

\vspace{-0.8em}
\begin{equation}
\label{eq:low_level_critic_loss}
\begin{aligned}
&\hspace*{-0.1cm}
\mathcal{L}(\theta^{Q,i}_l)=\mathop{\mathbb{E}}_{\textbf{x},\textbf{a},\textbf{r},\textbf{x}'\sim \mathcal{D}_l}\biggl[\Bigl(Q^i_l(\textbf{x}, a^1_l, ..., a^N_l)-y^i_l\Bigr)^2\biggr],
\\
&\hspace*{-0.1cm}
y^i_l=r^i_l+\gamma \bar{Q}^i_l(\textbf{x}',a'^1_l, ... ,a'^N_l)|_{a'^j_l=\bar{\mu}^j_l(s'^j_l)},
\end{aligned}
\end{equation}
where $\textbf{x}=\{s^1_l, s^2_l, ..., s^N_l\}$ collects the states of all low-level agents, and $\textbf{x}'$ contains all the next states. $\mathcal{D}_l$ is the replay buffer shared between low-level agents, storing the transition experience with tuples $(\textbf{x}, a^1_l, ... ,a^N_l, r^1_l, ... , r^N_l, \textbf{x}')$. $Q^i_l$ is the action-value function from the critic network of agent $c_i$.  $\bar{Q}^i_l$ and $\bar{\mu}^j_l$ are the target action-value function and the target policy, respectively.

\begin{figure}[t]
\graphicspath{{figs/}}
\begin{center}
\includegraphics[width=0.5\textwidth]{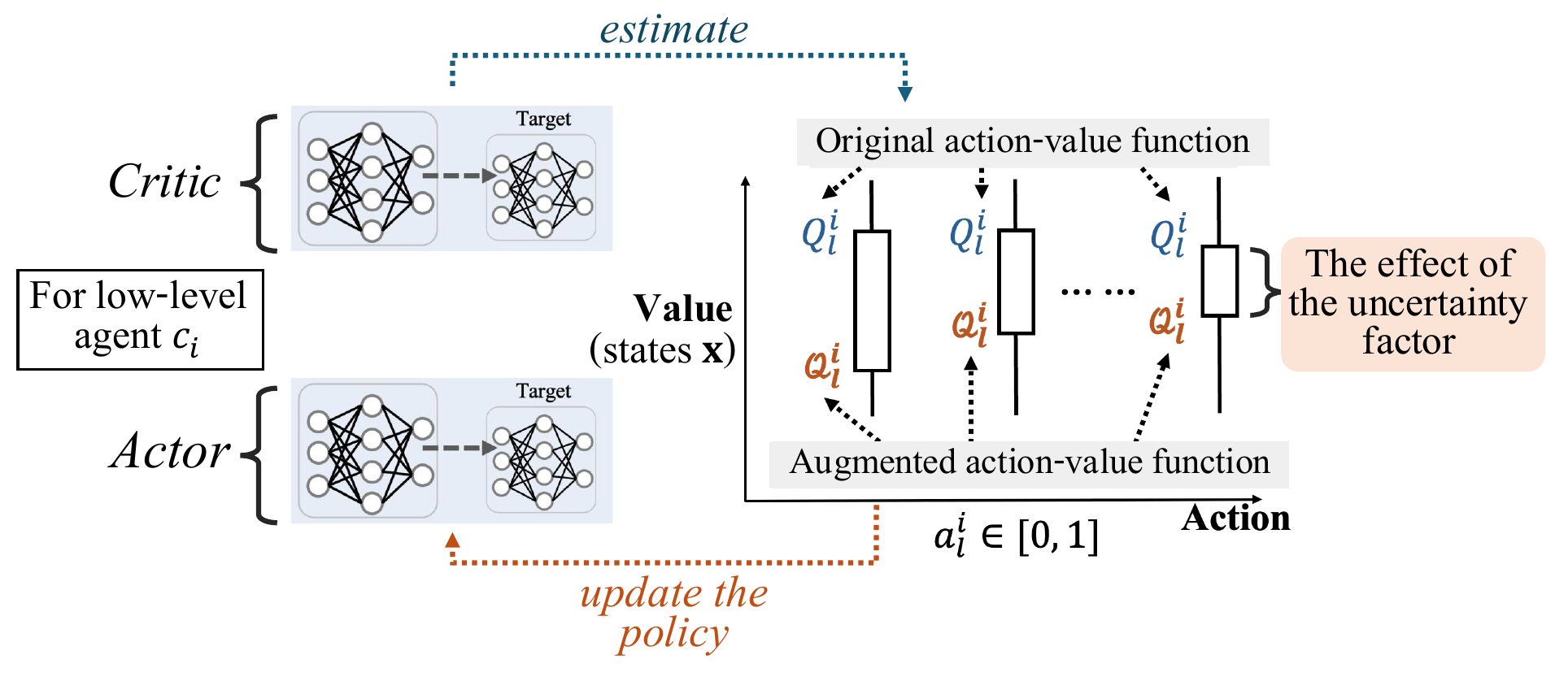}
\end{center}
\vspace{-1em}
\caption{The illustration of uncertainty-aware critic augmentation. For each low-level agent $c_i$, the action-value function is augmented based on the uncertainty factor (Eq.~\ref{eq:critic_aug}). This augmentation guides the updated policy to make decisions with explicit consideration of EV departure uncertainty.}
\vskip -1em
\label{fig:critic_aug}
\end{figure}

However, early departures of EVs occur occasionally, making the departure time uncertain and complicating the selection of the best low-level actions. To tackle this challenge, inspired by the Upper Confidence Bound (UCB) algorithm~\cite{auer2002finite}, which adjusts action-value estimations based on uncertainty (or confidence) and thereby affects the action decision, we propose a novel \textit{augmented action-value function}, $\mathcal{Q}^i_l$, for the \textbf{actor network} of each low-level agent $c_i$.

\noindent $\blacktriangleright$ \textbf{\underline{Uncertainty-Aware Critic Augmentation:}} Our idea is to discourage high-power discharging actions as the expected departure time approaches and the expected SoC remains unmet. This augmentation \ding{182} \textbf{integrates uncertainty into the actor network to guide decision-making}, while \ding{183} \textbf{shielding the critic network from the direct influence of uncertainties}, thereby preserving critic convergence and improving action selection. 
Consequently, we augment the action-value function based on the departure uncertainty after updating the critic network.

Our uncertainty-aware critic augmentation is illustrated in Figure~\ref{fig:critic_aug}, and the \textbf{augmented action-value function} $\mathcal{Q}^i_l$ is designed as follows:

\vspace{-1.5em}
\begin{equation}
\label{eq:critic_aug}
{\small
\begin{aligned}
&\hspace*{-0.3cm}
\mathcal{Q}^i_l(\textbf{x}, \textbf{a})=Q^i_l(\textbf{x}, a^1_l, ..., a^N_l)\cdot \biggl(1 -
\\
&\hspace*{1.8cm}
  \rho \cdot \underbrace{|\log_2 (a^i_l+\epsilon )| \cdot  \sqrt{\frac{\max(\Delta SoC^i_t,0)}{\Delta T^i_t} }}_{\text{the uncertainty factor}}\:\:\biggr)\Large \small \large,
\end{aligned}
}
\vspace{-0.3em}
\end{equation}
where $a^i_l \in [0,1]$ is the action chosen by low-level agent $c_i$. A smaller $a^i$ indicates a tendency for higher discharge power. $\Delta SoC^i_t$ represents the difference between the expected SoC $SoC^i_{\text{dep}}$ and the current SoC $SoC^i_t$ of the EV connected to low-level agent (pile) $c_i$. $\Delta T^i_t$ is the time difference between the current time $t$ and the planned departure time $t^i_{\text{dep}}$. $\rho \in \mathbb{R}$ is a fixed coefficient representing the impact of uncertainty, and $\epsilon$ is a small constant. Consequently, high power discharging action is penalized when the uncertainty factor increases.

Based on Eq.~\ref{eq:critic_aug}, the actor network of low-level agent $c_i$, with parameters $\theta^{\mu, i}_l$, updates the low-level policy $\mu^i_l$ with gradient descent as:

\vspace{-0.6em}
\begin{equation}
\label{eq:low_level_actor_loss}
{\small	
\nabla _{\theta^{\mu, i}_l}J(\mu^i_{l})=
\mathop{\mathbb{E}}_{\textbf{x},\textbf{a}\sim\mathcal{D}_l}\bigl[\nabla _{\theta^{\mu, i}_l}\mu^i_{l}(a^i_l|s^i_l)\nabla _{a^i_h}\mathcal{Q}^i_l(\textbf{x}, \textbf{a})|_{a^i_l=\mu^i_{l}(s^i_l)}\bigr].
}
\end{equation}

Similar to the high-level agent, each low-level agent is trained with mini-batch updates. At every training step, a low-level agent $c_i$ draws a mini-batch $\mathcal{B}_l$ of transitions $\bigl\{(\textbf{x}^{(b)}, a^{1,(b)}_l, ... ,a^{N, (b)}_l, r^{1,(b)}_l, ... , r^{N,(b)}_l, \textbf{x}'^{(b)})\bigr\}_{b=1}^{\mathcal{B}_l}$ from the low-level replay buffer $\mathcal{D}_l$. The corresponding critic and actor networks, parameterized by $\theta^{Q,i}_l$ and $\theta^{\mu, i}_l$, are then updated as described in Algorithm~\ref{algo:low_update}.

\begin{algorithm}[t]
\small
\caption{Low-Level Agent Update}
\label{algo:low_update}
\begin{algorithmic}[1]
\Procedure{UpdateLow}{$\mathcal{B}_l$, $\theta^{Q,i}_l$, $\theta^{\mu, i}_l$}
\State Update the critic network by minimizing:

$\mathcal{L}(\theta^{Q,i}_{l}) = \frac{1}{\mathcal{B}_l} \sum_b \Big(Q^{i}_l(\textbf{x}^{(b)}, a^{1, (b)}_l, ..., a^{N, (b)}_l) - y^{i, (b)}_l\Big)^2$
\\
\State Compute the augmented action-value function $\mathcal{Q}^i_l$ \Comment{Eq.~\ref{eq:critic_aug}}
\State Update the actor network based on $\mathcal{Q}^i_l$:

$\nabla_{\theta^{\mu, i}_l} J \approx \frac{1}{\mathcal{B}_l} \sum_b \nabla_{\theta^{\mu, i}_l} \mu^i_l(s^{i,(b)}_l) \nabla_{a^{i,(b)}_l} \mathcal{Q}^i_l(\textbf{x}^{(b)}, \textbf{a}^{(b)})$
\EndProcedure
\end{algorithmic}
\end{algorithm}

\noindent \textbf{Continuous actions and rewards for low-level agents.} In Eq.~\ref{eq:critic_aug}, the action $a^i_l$ is derived from the raw logits $g^i_l$ produced by the low-level policy $\mu^i_l$ of agent $c_i$. To further entangle low-level action space with the high-level discrete decision $a^{\text{disc}}_h$, the continuous action $a^{i}_l$ is formulated as follows:

\begin{equation}
\label{eq:low_level_action_value}
a^{i}_l = \left\{\begin{matrix}
\delta  + \delta\cdot \sigma (g^i_l) & \text{, if $a^{\text{disc}}_h=1$}  \\
\delta\cdot \sigma (g^i_l)  & \text{, if $a^{\text{disc}}_h=0$}
\end{matrix}\right.,
\end{equation}
where $\sigma(\cdot)$ is the sigmoid function, and $\delta$ is set to 0.5. In this setting, the action space of $a^{i}_l$ is constrained to $[0, 0.5)$ while the discrete high-level action is ``discharging'' ($a^{\text{disc}}_h=0$). In contrast, $a^{i}_l$ is limited to $[0.5, 1]$ when the high-level action is ``charging'' ($a^{\text{disc}}_h=1$).

Subsequently, the \textbf{optimal charging/discharging power} $\mathcal{P}^{\text{opt}}_{i,t}$ provided by low-level agent (pile) $c_i \in \mathcal{C}$ to the connected EV at time slot $t$ is estimated based on the continuous low-level action $a^{i}_l$, as defined below:
\begin{equation}
\label{eq:opt_charging_power}
\mathcal{P}^{\text{opt}}_{i,t}= a^{i}_l\cdot (\mathcal{P}^{\max}_{i, t} -\mathcal{P}^{\min}_{i,t}) + \mathcal{P}^{\min}_{i,t}.
\end{equation}

Based on the optimal power decision $\mathcal{P}^{\text{opt}}_{i,t}$ at each time slot, the reward $r^i_l$ for the low-level agent (pile) $c_i$ is formulated considering the charging cost at the current time slot $t$ and the difference between the current SoC and the expected SoC at the EV's departure connected to $c_i$, as defined below:

\begin{equation}
\label{eq:low_level_reward}
r^i_l=\omega\cdot (-\mathcal{P}^{\text{opt}}_{i,t}\cdot p_t)+ (-|SoC^i_{t}-SoC^i_{\text{dep}}|),
\end{equation}
where $p_t$ and $SoC^i_t$ are the energy price and the SoC of the EV at time slot $t$, respectively. $SoC^i_{\text{dep}}$ is the expected SoC at the EV's departure. The parameter $\omega$ reflects the impact of the current charging cost. This reward design encourages low-level agents to discharge EVs when the energy price is high, while also aiming to meet their expected SoC.

\subsection{Optimization for the Hierarchical Control.}
\label{subsec:optimal_strategy}

In Secs.~\ref{subsec:high-level} and \ref{subsec:low-level-uncertain-aug}, the critic and actor networks are optimized using mini-batches sampled from the replay buffers. To stabilize training, a soft target-network update is applied. Let $\theta_h=\{\theta^Q_h, \theta^{\mu}_h\}$ and $\theta^i_l=\{\theta^{Q,i}_l, \theta^{\mu,i}_l\}$ denote the parameters of the high-level and the \(i\)-th low-level networks, respectively,
and let $\theta_h'=\{\theta'^Q_h, \theta'^{\mu}_h\}$ and $\theta^{'i}_l=\{\theta'^{Q,i}_l, \theta'^{\mu,i}_l\}$ be their corresponding target network parameters.  
These target parameters are updated as follows:

\vspace{-0.6em}
\begin{equation}
\label{eq:soft_update}
     \theta'_h \leftarrow \tau \theta_h + (1 - \tau) \theta'_h, \quad \theta^{'i}_l \leftarrow \tau \theta^{i}_l + (1 - \tau) \theta^{'i}_l,
\end{equation}
where $\tau \in [0, 1]$ is the coefficient controlling the update rate.

The overall architecture of \emph{HUCA} is shown in Algorithm~\ref{alg:huca}. This design enables both high-level and low-level agents to learn optimal charging control policies by considering dynamic factors. Therefore, \emph{HUCA} provides more effective real-time charging control without requiring future information.

\begin{algorithm}[h]
\small
\caption{The \emph{HUCA} framework}
\label{alg:huca}
\begin{algorithmic}[1]
\Require The maximum learning episode $E$; charging piles $\mathcal{C}=\{c_1, c_2, ..., c_N\}$; EV set $\mathcal{V}_t$; EV charging information $\{\mathcal{I}^i|v_i\in \mathcal{V}_t\}$; power boundaries $\{\mathcal{P}^{\text{max}}_{i,t}, \mathcal{P}^{\text{min}}_{i,t}|v_i \in \mathcal{V}_t\}$.
\Ensure Optimal policies $\mu_h$ and $\{\mu_l^{\,i}\}_{i=1}^{N}$ with parameters $\theta^{\mu}_h$ and $\{\theta^{\mu,i}_l\}_{i=1}^{N}$

\State Initialize parameters
        $\theta^{\mu}_h,\theta^{Q}_h$ for the high-level agent and
        $\{\theta^{\mu,i}_l,\theta^{Q,i}_l\}_{i=1}^{N}$ for
        the low-level agents; 
        replay buffers $\mathcal{D}_h, \mathcal{D}_l$.
\For{episode = 1 to $E$}
    \State Receive initial state $s_h$, $\textbf{x} = \{s^i_l\}_{i=1}^N$
    \For{$t = 1$ to max-episode-length}
        \State {\fontfamily{qcr}\selectfont // Select actions}
        \State Select and compute action $a^{disc}_h$ 
        \For{each low-level agent $i = 1, ..., N$}
            \State Select and compute action $a^{i}_l$ based on $a^{disc}_h$ 
            \State Estimate optimal power $\mathcal{P}^{\text{opt}}_{i,t}$ 
        \EndFor
        \State {\fontfamily{qcr}\selectfont // Store transition experiences}
        \State Apply optimal powers $\{\mathcal{P}^{\text{opt}}_{i,t}\}_{i=1}^N$ to EVs and receive rewards $(r_h, \{r^i_l\}_{i=1}^N)$ 
        \State Observe next states $(s'_h, \textbf{x}' = \{s'^{i}_l\}_{i=1}^N)$

        \State Store $(s_h, a_h, r_h, s'_h)$ in $\mathcal{D}_h$
        \State Store $(\textbf{x}, a^1_l, ... ,a^N_l, r^1_l, ... , r^N_l, \textbf{x}')$ in $\mathcal{D}_l$
        \State Update current states
    \EndFor
    \State {\fontfamily{qcr}\selectfont // Update networks}
    \State Sample a mini-batch $\mathcal B_h$ from $\mathcal \mathcal{D}_h$
    \State \Call{UpdateHigh}{$\mathcal B_h$, $\theta^Q_h$, $\theta^{\mu}_h$} \Comment{Algorithm~\ref{algo:high_update}}
    \State Sample a mini-batch $\mathcal B_l$ from $\mathcal{D}_l$
    \For{each low-level agent $i = 1, ..., N$}
        \State \Call{UpdateLow}{$\mathcal B_l$, $\theta^{Q,i}_l$, $\theta^{\mu, i}_l$} 
        \Comment{Algorithm~\ref{algo:low_update}}
    \EndFor
    \State Soft target networks update 
\EndFor
\end{algorithmic}
\end{algorithm}

\section{Experimental Results}

\subsection{Dataset and Experimental Setup}

\subsubsection{\textbf{Dataset and Preprocessing.}}
For our workplace charging scenario, following datasets are used: \textbf{(i)} \underline{Building} \underline{electricity demands}: the CU-BEMS dataset~\cite{Pipattanasomporn2020} provides office building electricity consumption recorded at one-minute intervals from July 2018 to December 2019. \textbf{(ii)} \underline{Pricing} \underline{mechanism}: we use the ComEd APIs~\cite{ComEdHourlyPricing}, which provide real-time 5-minute electricity pricing data through its hourly pricing program. 
All data are aggregated into hourly intervals. The training period spans July 1 to August 31, 2018, with testing conducted in the following month.

\subsubsection{\textbf{Penalty Mechanism.}}
\label{appendix:penalty} The penalty charge is typically two or three times the basic capacity rate~\cite{rosado2020framework}. We design the penalty system based on the amount of electricity exceeding the contracted capacity (the upper bound of instant electricity consumption)~\cite{Fernndez2013CostOO}~\cite{rosado2020framework}, with detailed settings shown in Table~\ref{tab:penalty}. Let the maximum instant electricity load during the testing period be $L_{max}$. If $L_{max} < C_{contract}$, then the penalty is set to zero. Given a threshold coefficient $0 < \delta < 1$, the penalty is calculated by dividing the excess electricity into two tiers:

\begin{enumerate}[label=(\alph*)]
    \item Tier 1 (Up to 10\% Overload): For overloads up to 10\% of the contract capacity ($ 0 < L_{max} \leq \delta\cdot C_{contract} $), the penalty charge is set to $2\cdot R_{base}$.
    \item Tier 2 (Beyond 10\% Overload): For overloads exceeding 10\% of the contracted capacity ($ L_{max} > \delta\cdot C_{contract}$), the penalty charge is set to $ 3\cdot R_{base} $ for the amount that surpasses the 10\% threshold.
\end{enumerate}

Based on this setting, the total penalty cost is estimated as:

\vspace{-0.6em}
\begin{equation}
\begin{aligned}
&\hspace*{0cm}
\text{Penalty Cost} = 2\cdot R_{base}\cdot \min(L_{max}, \delta\cdot C_{contract}) 
\\
&\hspace*{2.5cm}
+ 3\cdot R_{base}\cdot \max(0, L_{max}-\delta\cdot C_{contract}).
\end{aligned}
\end{equation}

\begin{table}[b]
\small
\centering
\caption{Design of the penalty mechanism.}
\label{tab:penalty}
\begin{tabular}{lc}
\toprule[1.3pt]
\textbf{Parameter} & \textbf{Value} \\ 
\midrule
Contract Capacity ($C_{contract}$)       & 700~kW \\ 
Basic Capacity Rate  ($R_{base}$)             & \$15/kW/month \\
\midrule[1.3pt]
\textbf{Exceed Electricity Amount} & \textbf{Penalty Charge} \\ \hline
Up to 10\% over $C_{contract}$ & \( 2\cdot R_{base} \) \\ 
More than 10\% over $C_{contract}$ & \( 3\cdot R_{base} \) \\ 
\bottomrule[1.3pt]
\end{tabular}
\end{table}

\begin{table}[t]
\small
\centering
\caption{Random variables for EV information.}
\vspace{0.3em}
\begin{scriptsize}
\label{tab:charging_behavior}
\begin{tabular}{lcc}
\toprule[1.3pt]
\textbf{Information} & \textbf{Distribution} & \textbf{Boundaries} \\
\midrule
Arrival time  & $\mathcal{N}(9, 1^2)$   & $7 \leq t^i_{\text{arr}} \leq 12$ \\ 
Expected departure time & $\mathcal{N}(19, 1^2)$  & $16 \leq t^i_{\text{dep}} \leq 23$ \\ 
SoC upon arrival& $\mathcal{N}(0.4, 0.1^2)$ & $0.3 \leq SoC^i_{\text{arr}} \leq 0.6$ \\ 
Expected SoC at departure & $\mathcal{N}(0.8, 0.1^2)$ & $0.6 \leq SoC^i_{\text{dep}} \leq 0.9$ \\ 
\bottomrule[1.3pt]
\end{tabular}
\end{scriptsize}
\end{table}

\begin{table*}[t]
\centering
\caption{Performance comparison with different number of charging piles (CP). Except of ``OPT'', the best and second best results for ``Penalty Cost (USD)'' ($\downarrow$) and ``Total Cost (USD)'' ($\downarrow$) are in \textbf{bold} and \underline{underlined}, respectively. Cells in \colorbox{gray!25}{gray} highlight the results that are greater than or equal to the median value for ``SoC Maintenance (\%)'' ($\uparrow$), ``SoC Fulfillment (\%)'' ($\uparrow$) and ``User Satisfaction (\%)'' ($\uparrow$). Relative performance to \emph{HUCA} is given in parentheses as percentage differences (\textcolor{myred}{$\blacktriangle$better}/\textcolor{myblue}{$\triangledown$worse}).}
\renewcommand{\arraystretch}{1.2}
\setlength{\tabcolsep}{5pt}
\resizebox{\textwidth}{!}{
\begin{tabular}{lccccccc}
\toprule[1.3pt]
\multicolumn{8}{c}{\textbf{\textit{Certain Departure Scenario}}} \\ 
\midrule[1.3pt]
\multirow{2}{*}{\textbf{Metrics}} & \multirow{2}{*}{\textbf{\thead{CP\\Num.}}} & \multicolumn{6}{c}{\textbf{Methods}} \\
 & & \multicolumn{1}{c}{OPT} &  \multicolumn{1}{c}{DDPG}  & \multicolumn{1}{c}{IQL} & \multicolumn{1}{c}{VDN}  & \multicolumn{1}{c}{MADDPG} &\multicolumn{1}{c}{\textbf{\emph{HUCA} (ours)}} \\
\hline\hline
Penalty Cost ($\downarrow$) & \multirow{2}{*}{10} & 0.00$_{\textcolor{myred}{(\blacktriangle100.0\%)}}$ & 3682.85$_{\textcolor{myblue}{(\triangledown937.9\%)}}$ & 2516.84$_{\textcolor{myblue}{(\triangledown609.3\%)}}$ & 522.28$_{\textcolor{myblue}{(\triangledown47.2\%)}}$ & \underline{354.84}$_{\textcolor{myblue}{(\triangledown0.0\%)}}$ & \textbf{354.82} \\
Total Cost ($\downarrow$) &  & \multicolumn{1}{c}{5805.28$_{\textcolor{myred}{(\blacktriangle6.6\%)}}$}  & 9563.31$_{\textcolor{myblue}{(\triangledown53.9\%)}}$ & 8426.36$_{\textcolor{myblue}{(\triangledown35.6\%)}}$ & 6431.31$_{\textcolor{myblue}{(\triangledown3.5\%)}}$  &  \underline{6220.33}$_{\textcolor{myblue}{(\triangledown0.1\%)}}$ & \textbf{6215.42} \\
\midrule
Penalty Cost ($\downarrow$) & \multirow{2}{*}{20} & 0.00$_{\textcolor{myred}{(\blacktriangle100.0\%)}}$ & 7818.52$_{\textcolor{myblue}{(\triangledown375.7\%)}}$ & 5643.15$_{\textcolor{myblue}{(\triangledown243.4\%)}}$ & 2271.32$_{\textcolor{myblue}{(\triangledown38.2\%)}}$ & \underline{1869.38}$_{\textcolor{myblue}{(\triangledown13.7\%)}}$ & \textbf{1643.47} \\
Total Cost ($\downarrow$) &  & 5860.05$_{\textcolor{myred}{(\blacktriangle24.2\%)}}$ & 13929.11$_{\textcolor{myblue}{(\triangledown80.2\%)}}$ & 11798.68$_{\textcolor{myblue}{(\triangledown52.7\%)}}$ & 8413.61$_{\textcolor{myblue}{(\triangledown8.9\%)}}$ & \underline{7952.69}$_{\textcolor{myblue}{(\triangledown2.9\%)}}$ & \textbf{7728.57}\\ 
\toprule[1.3pt]
\midrule[1.3pt]
\multicolumn{8}{@{}c@{}}{\textbf{\textit{Uncertain Departure Scenario}}}\\
\midrule[1.3pt]
\multirow{2}{*}{\textbf{Metrics}} & \multirow{2}{*}{\textbf{\thead{CP\\Num.}}} & 
  \multicolumn{6}{c}{\textbf{Methods}} \\
 & & OPT & DDPG & IQL & VDN & MADDPG & \textbf{\emph{HUCA} (ours)} \\
\hline\hline
Penalty Cost ($\downarrow$) & \multirow{4}{*}{10} & 0.00$_{\textcolor{myred}{(\blacktriangle100.0\%)}}$ & 4671.74$_{\textcolor{myblue}{(\triangledown1212.9\%)}}$ & 1742.35$_{\textcolor{myblue}{(\triangledown389.6\%)}}$ & 529.18$_{\textcolor{myblue}{(\triangledown48.7\%)}}$ & \underline{448.52}$_{\textcolor{myblue}{(\triangledown26.0\%)}}$ & \textbf{355.84} \\
Total Cost ($\downarrow$) & & 5691.96$_{\textcolor{myred}{(\blacktriangle5.0\%)}}$ & 10398.77$_{\textcolor{myblue}{(\triangledown73.5\%)}}$ & 7479.04$_{\textcolor{myblue}{(\triangledown24.8\%)}}$ & 6214.53$_{\textcolor{myblue}{(\triangledown3.7\%)}}$ & \underline{6087.48}$_{\textcolor{myblue}{(\triangledown1.6\%)}}$ & \textbf{5993.78} \\
\cdashline{1-1}\cdashline{3-8}
SoC Fulfillment ($\uparrow$) & & 53.77$_{\textcolor{myred}{(\blacktriangle31.9\%)}}$  & \cellcolor{gray!25}54.05$_{\textcolor{myred}{(\blacktriangle32.6\%)}}$ & \cellcolor{gray!25}46.04$_{\textcolor{myred}{(\blacktriangle12.9\%)}}$ & 40.34$_{\textcolor{myblue}{(\triangledown1.1\%)}}$ & 27.10$_{\textcolor{myblue}{(\triangledown33.5\%)}}$ & \cellcolor{gray!25}40.77  \\
SoC Maintenance ($\uparrow$) & & 42.52$_{\textcolor{myblue}{(\triangledown6.3\%)}}$   & -10.58$_{\textcolor{myblue}{(\triangledown123.3\%)}}$ & \cellcolor{gray!25}129.01$_{\textcolor{myred}{(\blacktriangle184.2\%)}}$ & -120.61$_{\textcolor{myblue}{(\triangledown365.7\%)}}$ & \cellcolor{gray!25}11.87$_{\textcolor{myblue}{(\triangledown73.9\%)}}$ & \cellcolor{gray!25}45.40 \\
User Satisfaction ($\uparrow$) && 48.14$_{\textcolor{myred}{(\blacktriangle11.7\%)}}$	 & \cellcolor{gray!25}21.73$_{\textcolor{myblue}{(\triangledown49.6\%)}}$ &	\cellcolor{gray!25}87.52$_{\textcolor{myred}{(\blacktriangle103.2\%)}}$ &	-40.13$_{\textcolor{myblue}{(\triangledown193.2\%)}}$ & 19.48$_{\textcolor{myblue}{(\triangledown54.8\%)}}$ & \cellcolor{gray!25}43.08\\
\midrule
Penalty Cost ($\downarrow$) & \multirow{4}{*}{20} & 0.00$_{\textcolor{myred}{(\blacktriangle100.0\%)}}$ & 5010.81$_{\textcolor{myblue}{(\triangledown1311.6\%)}}$ & 2002.12$_{\textcolor{myblue}{(\triangledown464.1\%)}}$ & 966.51$_{\textcolor{myblue}{(\triangledown172.3\%)}}$ & \underline{931.01}$_{\textcolor{myblue}{(\triangledown162.3\%)}}$ &  \textbf{354.90}\\
Total Cost ($\downarrow$) &  & 5760.05$_{\textcolor{myred}{(\blacktriangle4.2\%)}}$ & 10823.07$_{\textcolor{myblue}{(\triangledown80.0\%)}}$ & 7815.99$_{\textcolor{myblue}{(\triangledown30.0\%)}}$ & 6721.82$_{\textcolor{myblue}{(\triangledown11.8\%)}}$ & \underline{6562.80}$_{\textcolor{myblue}{(\triangledown9.1\%)}}$ &  \textbf{6013.52} \\
\cdashline{1-1}\cdashline{3-8}
SoC Fulfillment ($\uparrow$) &  & 56.06$_{\textcolor{myred}{(\blacktriangle32.1\%)}}$ & \cellcolor{gray!25}55.27$_{\textcolor{myred}{(\blacktriangle30.2\%)}}$ & \cellcolor{gray!25}47.94$_{\textcolor{myred}{(\blacktriangle13.0\%)}}$ &  \cellcolor{gray!25}48.88$_{\textcolor{myred}{(\blacktriangle15.2\%)}}$ & 40.34$_{\textcolor{myblue}{(\triangledown4.9\%)}}$ & 42.44\\
SoC Maintenance ($\uparrow$)  & & 32.77$_{\textcolor{myblue}{(\triangledown73.7\%)}}$ & \cellcolor{gray!25}12.56$_{\textcolor{myblue}{(\triangledown89.9\%)}}$ & 0.90$_{\textcolor{myblue}{(\triangledown99.3\%)}}$ & \cellcolor{gray!25}91.96$_{\textcolor{myblue}{(\triangledown26.1\%)}}$ & -1.16$_{\textcolor{myblue}{(\triangledown100.9\%)}}$ & \cellcolor{gray!25}124.41 \\
User Satisfaction ($\uparrow$) && 44.41$_{\textcolor{myblue}{(\triangledown46.8\%)}}$ & \cellcolor{gray!25}33.91$_{\textcolor{myblue}{(\triangledown59.4\%)}}$ & 24.42$_{\textcolor{myblue}{(\triangledown70.7\%)}}$ & \cellcolor{gray!25}70.42$_{\textcolor{myblue}{(\triangledown15.6\%)}}$ & 19.59$_{\textcolor{myblue}{(\triangledown76.5\%)}}$ & \cellcolor{gray!25}83.42 \\
\bottomrule[1.3pt]
\end{tabular}}
\label{tb:baseline_compared}
\end{table*}

\subsubsection{\textbf{Simulation of EV Behaviors.}} The charging information $\mathcal{I}^i = (t^i_\text{arr}, t^i_\text{dep}, SoC^i_\text{arr}, SoC^i_\text{dep}, C^i)$ (Def. 2) of EV $v_i$ is modeled using normal distributions by following~\cite{li2023constrained}. Table~\ref{tab:charging_behavior} shows the setting of random variables and $C^i$ is set to 60 kWh. Two departure scenarios for determining the \underline{actual departure time} $\widehat{t}^i_{dep}$ of EV $v_i$ are examined:

\begin{enumerate}[label=(\alph*)]
    \item \textbf{Certain departure scenario}: $\widehat{t}^i_{dep}$ is the same as the expected departure time $t^i_{dep}$.
    \item \textbf{Uncertain departure scenario}: $\widehat{t}^i_{dep}$  is randomly sampled earlier than the expected departure time, $\widehat{t}^i_{dep}\in [1, t^i_{dep})$.
\end{enumerate}

\subsubsection{\textbf{Baselines.}} We compare our \emph{HUCA} against several baselines, including an oracle with full knowledge of future information, the single-agent model(\textbf{DDPG}~\cite{lillicrap2015continuous}) and multi-agent reinforcement learning models (\textbf{IQL}~\cite{Tan1997MultiAgentRL}, \textbf{VDN}\cite{Sunehag2017ValueDecompositionNF}, and \textbf{MADDPG}~\cite{lowe2017multi}), as described below:

\begin{itemize}[leftmargin=*]
    \item \textbf{OPT}: Assumes complete knowledge of future information, including all EV charging schedules, building energy demands, and dynamic electricity prices. This baseline uses a linear programming optimization model to solve the charging control problem.
    \item \textbf{DDPG (Deep Deterministic Policy Gradient)}~\cite{lillicrap2015continuous}: An actor-critic algorithm designed for continuous control in single-agent environments.
    \item \textbf{IQL (Independent Q-Learning)} \cite{Tan1997MultiAgentRL}: A decentralized multi-agent reinforcement learning approach where each agent independently learns its own Q-function.
    \item \textbf{VDN (Value Decomposition Networks)}~\cite{Sunehag2017ValueDecompositionNF}: A centralized value-based reinforcement learning method used in multi-agent systems.
    \item \textbf{MADDPG (Multi-Agent Deep Deterministic Policy Gradient)} \cite{lowe2017multi}: Extends DDPG to multi-agent settings by incorporating centralized training and decentralized execution.
\end{itemize}

Note that for all models (except for OPT), the states, actions, and rewards are configured in the same manner as for our low-level agents, but without incorporating the information provided by the high-level agent network.

\subsubsection{\textbf{Evaluation Metrics.}} Following metrics are used to evaluate performance:
\begin{itemize}[leftmargin=*]
    \item \textbf{Penalty Cost (USD)} measures the cost of exceeding the contracted capacity~\cite{Fernndez2013CostOO}, which multiplies the penalty rate with the peak amount of electricity. In accordance with~\cite{rosado2020framework}, the penalty mechanism is designed as detailed in Table~\ref{tab:penalty}.
    \item \textbf{Total Cost (USD)} is the sum of the basic electricity cost (electricity load $L_t$ multiplied by dynamic pricing $p_t$) and the penalty cost over the testing data.
\end{itemize}

In the \textbf{certain departure scenario}, \textit{all EV charging demands are met} since all baselines are subject to the charging power limitations (Sec.~\ref{subsec:definition}, Def. 3) and no unexpected EV departures occur. In the \textbf{uncertain departure scenario}, how well the charging demands are fulfilled when unexpected EV departures happen needs examinations. Given $\widehat{SoC}^i_{dep}$ and $SoC^i_{dep}$ denoting the actual and expected SoC of EV $v_i$ at the time of departure, and $SoC^i_{arr}$ is the SoC of $v_i$ upon arrival at the parking lot, the metrics are defined as follows:

\begin{itemize}[leftmargin=*]
    \item \textbf{SoC Fulfillment (\%)} evaluates how well the expected SoC of EV $v_i$ is fulfilled, defined as: $\widehat{SoC}^i_{\text{dep}}/SoC^i_{\text{dep}}$.
    \item \textbf{SoC Maintenance (\%)} checks the fulfillment status at the midpoint ($\widehat{SoC}^i_{\text{mid}}$) between the arrival and the actual departure time, verifying whether the SoC is consistently maintained under the charging control. A drop below the arrival SoC would result in extra expenses billed to users. The metric is defined as: $(\widehat{SoC}^i_{\text{mid}}-SoC^i_{\text{arr}})/(\widehat{SoC}^i_{\text{dep}}-SoC^i_{\text{arr}})$, where $\widehat{SoC}^i_{\text{mid}}$ denotes the SoC at the midpoint. 
    \item \textbf{User Satisfaction (\%)} represents the average of SoC fulfillment and SoC maintenance, indicating EV users’ satisfaction with the charging performance. 
\end{itemize}
The performance of these three metrics is reported as the average for all EVs per parking session.

\subsubsection{\textbf{Hyperparamter Settings.}}
The charging capacity ($\mathcal{P}^{\max}$) is set to 150 (kW), the EV battery capacity $C^i$ (Def. 2) is 60 (kwh), and the charging efficiency is 95\%. For training our \emph{HUCA} model, we use 1,500 episodes, with a replay buffer capacity of 30,000 for each low-level agent, and a batch size of 1,024. The penalty coefficient $\varphi$ (Eq.~\ref{eq:high_level_reward}) weight factors $\kappa$ (Eq.~\ref{eq:high_level_reward}), $\omega$ (Eq.~\ref{eq:low_level_reward}), and the importance of the uncertainty term $\rho$ (Eq.~\ref{eq:critic_aug}) are set to 0.1, 0.1, 0.5, and 10, respectively.

\subsection{Comparison Results.}

As shown in Table~\ref{tb:baseline_compared}, except for the "OPT" method (best-case result), \textbf{\emph{HUCA} achieves the lowest penalty and total cost without relying on future information} (such as future energy prices, EV charging requests, and actual EV departure times) for charging control across both departure scenarios. In addition, its total cost is comparable to OPT, manifesting its effectiveness in minimizing costs. While the SoC fulfillment, maintenance, and user satisfaction scores of \emph{HUCA} are not the best, its performance \textbf{exceeds the median of the baselines in most cases} (highlighted in gray). In contrast, although DDPG and IQL achieve higher SoC fulfillment under different numbers of charging piles, they incur \textbf{substantially higher total costs than \emph{HUCA} (by 20-80\%)}. 

Combining the above observations, \emph{HUCA} actually \textbf{strikes the best balance between the trade-off of electricity costs and fulfilling user demands} by introducing the hierarchical control and the uncertainty-aware critic augmentation.

Among multi-agent models, VDN and MADDPG come closest to \emph{HUCA} in terms of cost. However, user satisfaction scores of MADDPG are much lower than those of \emph{HUCA}. While VDN achieves higher SoC fulfillment than \emph{HUCA} with 20 charging piles, its maintenance scores are significantly lower. In some cases, the maintenance scores of MADDPG and VDN fall into \underline{negative values}, indicating that EVs primarily serve as power providers and return with lower SoCs than at arrival, leading to additional expenses for users.

\begin{table}[t]
\footnotesize
\centering
\caption{Ablation study results for the uncertain departure scenario. (SoC fulfillment and maintenance scores are abbreviated as ``Ful.'' and ``Main.'', respectively.) Relative performance to \emph{HUCA}'s full model is given in (\textcolor{myred}{$\blacktriangle$better}/\textcolor{myblue}{$\triangledown$worse}).}
\setlength{\tabcolsep}{5pt}
\resizebox{0.5\textwidth}{!}{
\begin{tabular}{clccc}
\toprule[1.3pt]
\multicolumn{5}{c}{\textbf{\textit{Uncertain Departure Scenario}}} \\ 
\midrule[1.3pt]
\textbf{CP} & \multirow{1}{*}{\textbf{Method}} &  Penalty ($\downarrow$) & Ful. ($\uparrow$) & Main. ($\uparrow$)\\
\midrule
\multirow{4}{*}{10} & \textbf{\emph{HUCA}} & 355.84 & 40.77 & 45.40 \\
& w/o C.A. & 2015.62$_{\textcolor{myblue}{(\triangledown466.4\%)}}$ & 62.49$_{\textcolor{myred}{(\blacktriangle53.3\%)}}$ & 124.01$_{\textcolor{myred}{(\blacktriangle173.1\%)}}$  \\
& w/o H. & 2707.71$_{\textcolor{myblue}{(\triangledown660.9\%)}}$ & 63.10$_{\textcolor{myred}{(\blacktriangle54.8\%)}}$ & 31.91$_{\textcolor{myblue}{(\triangledown29.7\%)}}$ \\
& w/o Either & 723.46$_{\textcolor{myblue}{(\triangledown103.3\%)}}$& 44.53$_{\textcolor{myred}{(\blacktriangle9.2\%)}}$ & -12.53$_{\textcolor{myblue}{(\triangledown127.6\%)}}$  \\
\toprule[1.3pt]
\textbf{CP} & \multirow{1}{*}{\textbf{Method}} & Penalty ($\downarrow$) & Ful. ($\uparrow$) & Main. ($\uparrow$) \\
\midrule
\multirow{4}{*}{20} & \textbf{\emph{HUCA}} & 354.90 & 42.44 &  124.41  \\
& w/o C.A. & 9362.87$_{\textcolor{myblue}{(\triangledown2538.2\%)}}$ & 69.73$_{\textcolor{myred}{(\blacktriangle64.3\%)}}$ & 209.70$_{\textcolor{myred}{(\blacktriangle68.6\%)}}$  \\
& w/o H. & 3461.37$_{\textcolor{myblue}{(\triangledown875.2\%)}}$ & 58.61$_{\textcolor{myred}{(\blacktriangle38.1\%)}}$ & 7.69$_{\textcolor{myblue}{(\triangledown93.8\%)}}$  \\
& w/o Either  & 618.46$_{\textcolor{myblue}{(\triangledown74.3\%)}}$ & 53.26$_{\textcolor{myred}{(\blacktriangle25.5\%)}}$ & 43.02$_{\textcolor{myblue}{(\triangledown65.4\%)}}$  \\
\bottomrule[1.3pt]
\vspace{-6mm}
\end{tabular}}
\label{tb:ablation}
\end{table}

\subsection{Ablation Study and Uncertainty Impact.}
\begin{figure}[t]
\graphicspath{{figs/}}
\begin{center}
\includegraphics[width=0.45\textwidth]{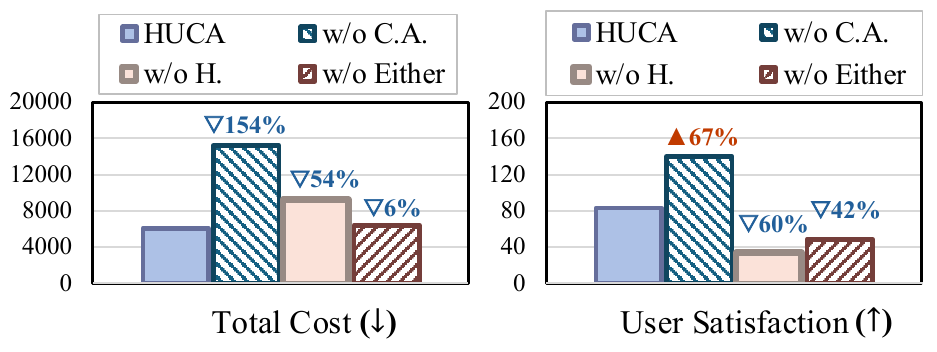}
\end{center}
\vspace{-1.5em}
\caption{Ablation study on total cost and user satisfaction (CP=20).}
\vspace{-4mm}
\label{fig:ablation_pile20}
\end{figure}

We conduct an ablation study by removing the uncertainty-aware critic augmentation (\textbf{w/o C.A.}), the high-level agent (\textbf{w/o H.}), and both of them (\textbf{w/o Either}). The results are shown in Table~\ref{tb:ablation} and Figure~\ref{fig:ablation_pile20}. In Table~\ref{tb:ablation}, removing either the critic augmentation or the high-level agent increases SoC fulfillment and maintenance scores (improvements of about 9–173\%). However, this also results in a much larger rise in penalty costs (worsening by about 74–2538\%). \textbf{The increase in penalty costs far outweighs the gains in SoC fulfillment and maintenance} when removing each component. On the other hand, removing both components results in a slight increase in costs but causes a significant drop in SoC maintenance scores. Figure~\ref{fig:ablation_pile20} also shows that with more charging piles (CP=20), removing critic augmentation increases user satisfaction but also raises the total cost substantially (worsening by 154\%) compared to the full \emph{HUCA} model.

These findings indicate that the hierarchical structure and critic augmentation in \emph{HUCA} complement each other, and their combined use is essential for achieving the lowest costs while balancing energy supply between the building and EVs.

\begin{figure}[b]
\graphicspath{{figs/}}
\begin{center}
\includegraphics[width=0.45\textwidth]{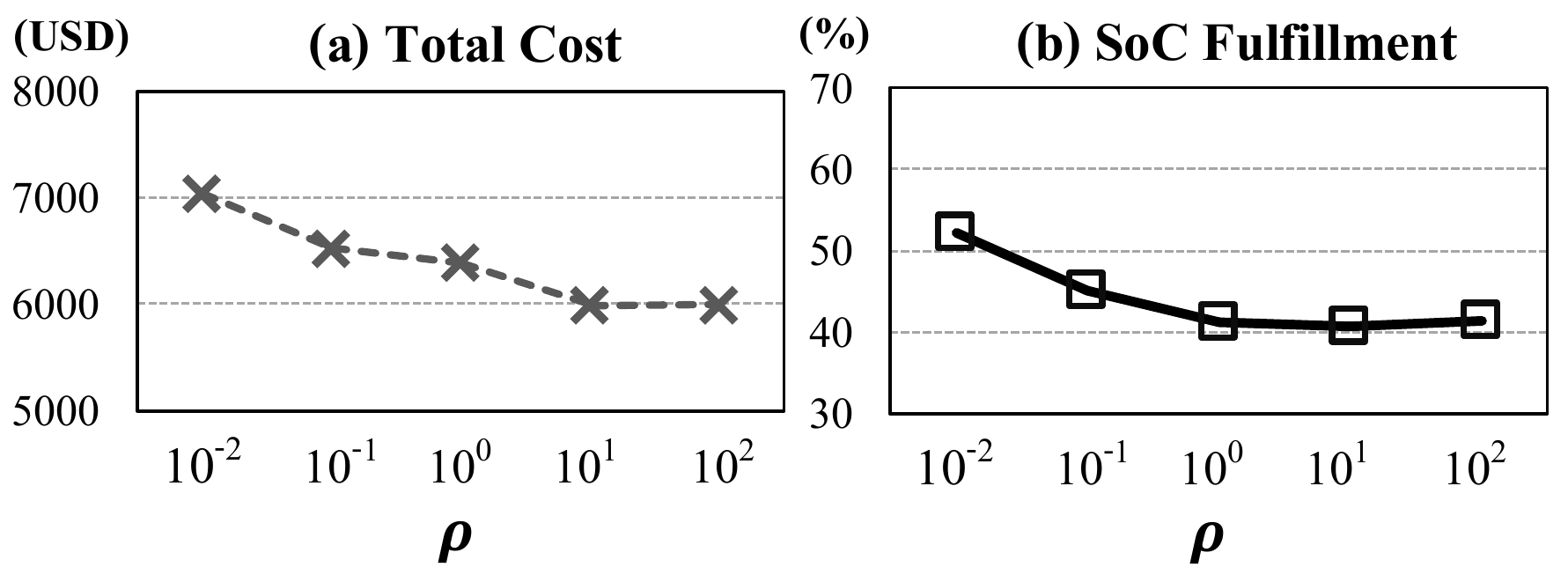}
\end{center}
\vspace{-1.5em}
\caption{Performance of \emph{HUCA} with different $\rho$ values with 10 piles in the uncertain departure scenario.}
\label{fig:uncertain_weight}
\end{figure}

Furthermore, to assess the impact of the uncertainty term in \emph{HUCA} for handling uncertain departures, Figure~\ref{fig:uncertain_weight} compares total cost and the SoC fulfillment under various uncertainty coefficient $\rho$ (Eq.~\ref{eq:critic_aug}). As $\rho$ increases, total cost decreases significantly. Meanwhile, setting $\rho$ to its lowest value ($10^{-2}$) improves SoC fulfillment but comes at the cost of significantly higher total costs. These findings suggest that $\rho$ should not be set too low to ensure a better trade-off between cost and user charging demand. Properly tuning $\rho$ can further improve this balance, enhancing the overall effectiveness of \emph{HUCA}.

\begin{figure}[t]
\graphicspath{{figs/}}
\begin{center}
\includegraphics[width=0.49\textwidth]{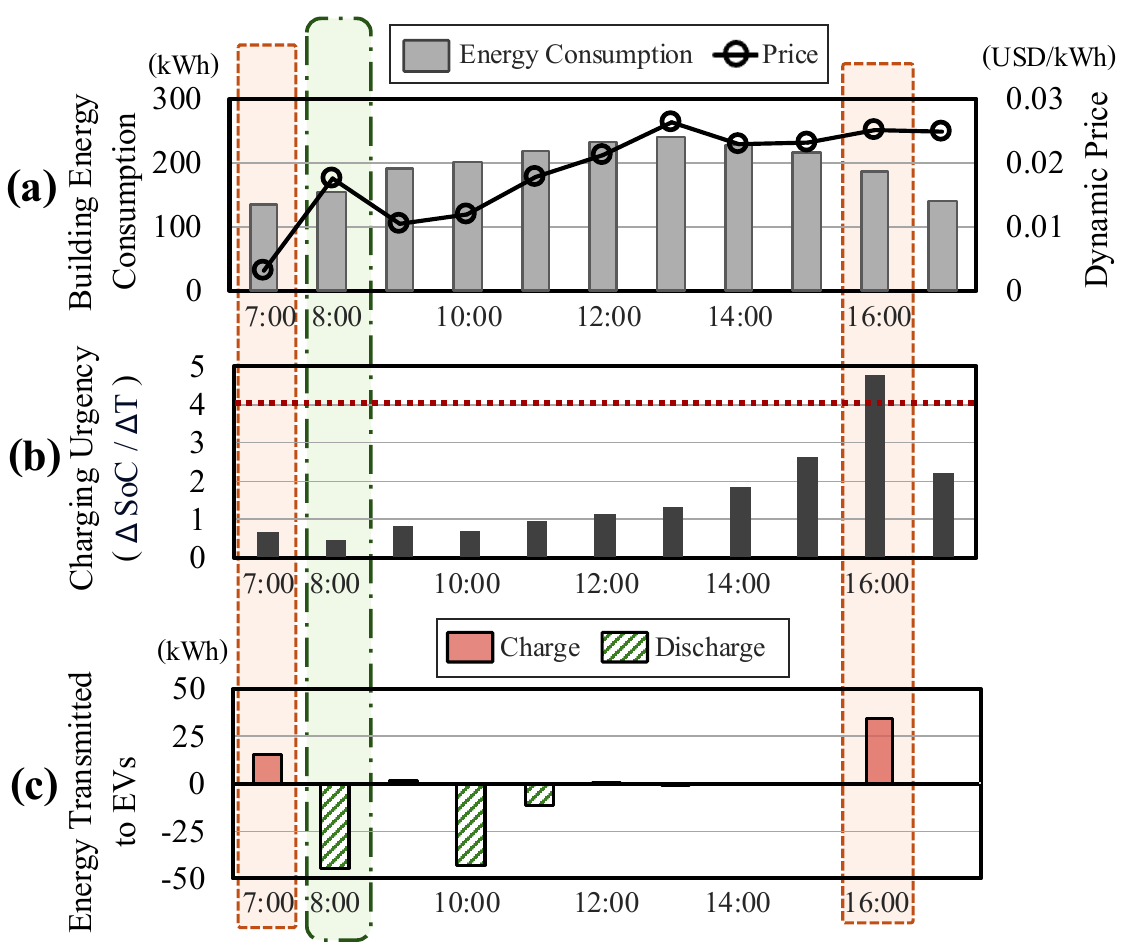}
\end{center}
\vskip -1.0em
\caption{Details of the charging control for a day from 7:00 to 17:00. (a) shows building electricity demands and dynamic pricing; (b) visualizes average charging urgency, with $\Delta SoC$ as the difference between current and expected SoC, and $\Delta T$ as the remaining time until planned departure; (c) presents final control decisions.}
\vspace{-4mm}
\label{fig:case_study}
\end{figure}

\subsection{Case Study.}
To examine how \emph{HUCA} balances energy supply between the building and EVs, Figure~\ref{fig:case_study} visualizes the control decisions over a single day for all EVs, which can be discussed in terms of charging and discharging behaviors:

\noindent\textbf{\underline{Charging Behaviors}.} Two significant EV charging behaviors appear at 7:00 and 16:00 (red rectangle) but with different reasons. At 7:00, both electricity demand and prices for the building are minimal (Figure~\ref{fig:case_study}(a)), therefore \emph{HUCA} focuses on charging the EVs to reach their target SoC. At 16:00, even though there is a high demand for building energy and elevated prices, which typically would lead to transferring energy from the EVs to the building to reduce costs, \emph{HUCA} opts to charge the EVs because of their high charging urgency (Figure~\ref{fig:case_study}(b)).

\noindent\textbf{\underline{Discharging Behaviors}.} A similar situation occurs at 8:00 (green rectangle); however, since the charging urgency is relatively low at this time, \emph{HUCA} discharges the EVs to serve the high-usage building while simultaneously avoiding high-priced electricity usage.

These results manifest that \emph{HUCA} effectively and dynamically balances energy supply between the building and EVs based on real-time information.

\section{Conclusion and Future Work}

In this paper, we propose \emph{HUCA}, a novel hierarchical multi-agent framework designed for real-time charging control. To address practical dynamics and limitations, \emph{HUCA} integrates hierarchical control and uncertainty-aware critic augmentation to adapt to dynamic factors, optimize charging decisions, and account for departure uncertainties. Experiments show that \emph{HUCA} outperforms existing methods in both certain and uncertain departure scenarios. A case study further illustrates that \emph{HUCA} effectively balances energy supply between the building and EVs using real-time information. 

One line of the future work is to account for user willingness when adjusting charging preferences to supply energy to the building, e.g., with both user willingness and buy-and-sell energy behaviors in \textit{HUCA}. Incorporating them can help improve the estimation of the confidence bound for the action-value function, leading to a more efficient charging control.

\bibliographystyle{plain} 
\bibliography{reference}

\end{document}